%% file: conference_041818.tex
\documentclass[conference,10pt]{IEEEtran}

\IEEEoverridecommandlockouts
\usepackage{authblk}

\usepackage[font=small,labelfont=bf, font=bf]{caption}
\def\tablename{Table}
\renewcommand{\thetable}{\arabic{table}}
\usepackage{multirow}
\usepackage{float}

\usepackage{verbatim}

\usepackage{cite}
\usepackage{graphicx}
\graphicspath{{images/}}

\usepackage{hyperref}

\begin{document}

\title{Platform Independent Software Analysis for Near Memory Computing}

\author{Stefano Corda$^{1, 2}$, Gagandeep Singh$^{1, 2}$, Ahsan Javed Awan$^3$, Roel Jordans$^1$, Henk Corporaal$^1$\\

\vspace{-0.4cm} \normalsize $^1$Eindhoven University of Technology   \hspace{1cm}$^2$IBM Research - Zurich \hspace{1cm} $^3$Ericsson Research\\\\

\{s.corda, g.singh, r.jordans, h.corporaal\}@tue.nl, ahsan.javed.awan@ericsson.com}

\maketitle

\input{sections/abstract.tex}

\begin{IEEEkeywords}
LLVM, NMC, Spatial Locality, Memory Entropy, Data-Level Parallelism
\end{IEEEkeywords}

%%%%%%%%%%%%%%%%%%%%%%%%%%%%%%%%%%%%%%%%%%%%%%%%%%%%%%%%%%%%%%%%%%%%%%%%%%%%%%%%%%%%%%%%%%
\section{Introduction}
\label{sec:introduction}
\input{sections/introduction.tex}

%%%%%%%%%%%%%%%%%%%%%%%%%%%%%%%%%%%%%%%%%%%%%%%%%%%%%%%%%%%%%%%%%%%%%%%%%%%%%%%%%%%%%%%%%%
\section{PISA-NMC}
\label{sec:backgrounds}
\input{sections/back_pisa.tex}
%%%%%%%%%%%%%%%%%%%%%%%%%%%%%%%%%%%%%%%%%%%%%%%%%%%%%%%%%%%%%%%%%%%%%%%%%%%%%%%%%%%%%%%%%%

%%%%%%%%%%%%%%%%%%%%%%%%%%%%%%%%%%%%%%%%%%%%%%%%%%%%%%%%%%%%%%%%%%%%%%%%%%%%%%%%%%%%%%%%%%
\section{Methodology}
\label{sec:methodology}
\input{sections/methodology.tex}

%%%%%%%%%%%%%%%%%%%%%%%%%%%%%%%%%%%%%%%%%%%%%%%%%%%%%%%%%%%%%%%%%%%%%%%%%%%%%%%%%%%%%%%%%%

%%%%%%%%%%%%%%%%%%%%%%%%%%%%%%%%%%%%%%%%%%%%%%%%%%%%%%%%%%%%%%%%%%%%%%%%%%%%%%%%%%%%%%%%%%
\section{Results}
\label{sec:results}
\input{sections/results.tex}

\section{Related Work}
\label{sec:relatedworks}
\input{sections/relatedworks.tex}

%%%%%%%%%%%%%%%%%%%%%%%%%%%%%%%%%%%%%%%%%%%%%%%%%%%%%%%%%%%%%%%%%%%%%%%%%%%%%%%%%%%%%%%%%%

%%%%%%%%%%%%%%%%%%%%%%%%%%%%%%%%%%%%%%%%%%%%%%%%%%%%%%%%%%%%%%%%%%%%%%%%%%%%%%%%%%%%%%%%%%
\section{Conclusion}
\label{sec:conclusions}
\input{sections/conclusions.tex}

%%%%%%%%%%%%%%%%%%%%%%%%%%%%%%%%%%%%%%%%%%%%%%%%%%%%%%%%%%%%%%%%%%%%%%%%%%%%%%%%%%%%%%%%%%

%%%%%%%%%%%%%%%%%%%%%%%%%%%%%%%%%%%%%%%%%%%%%%%%%%%%%%%%%%%%%%%%%%%%%%%%%%%%%%%%%%%%%%%%%%
\section*{Acknowledgments}
\input{sections/acks}
%%%%%%%%%%%%%%%%%%%%%%%%%%%%%%%%%%%%%%%%%%%%%%%%%%%%%%%%%%%%%%%%%%%%%%%%%%%%%%%%%%%%%%%%%%

\bibliographystyle{IEEEtran}
\bibliography{IEEEabrv,refshort}

\end{document}

%% file: sections/abstract.tex
\begin{abstract}
Near-memory Computing (NMC) promises improved performance for the applications that can exploit the features of emerging memory technologies such as 3D-stacked memory. However, it is not trivial to find such applications and specialized tools are needed to identify them. In this paper, we present PISA-NMC, which extends a state-of-the-art hardware agnostic profiling tool with metrics concerning memory and parallelism, which are relevant for NMC. The metrics include memory entropy, spatial locality, data-level, and basic-block-level parallelism. By profiling a set of representative applications and correlating the metrics with the application's performance on a simulated NMC system, we verify the importance of those metrics. Finally, we demonstrate which metrics are useful in identifying applications suitable for NMC architectures.
\end{abstract}

%% file: sections/introduction.tex
%With the demise of Dennard scaling and slowing of Moore's law, computing performance is hitting a plateau \cite{Esmaeilzadeh:2011:DSE:2024723.2000108}. The improvements in memory and processor technology have grown at a different speed, which is infamously termed as the memory wall \cite{Wulf:1995:HMW:216585.216588}. This technological trend has resulted in a situation where the data transfers to and from memory are slower than the processing capabilities of processors. This extreme transition has introduced several challenges in the design of future computing architecture.
 
%Additionally, the era of big-data---where data is generated in a missive amount and from multiple domains---has created a demand for novel memory-centric designs rather than conventional compute-centric designs~\cite{overviewpaper}. In conventional systems, a deep layer of cache memories sits in between the core and the main memory to reduce the gap between memory and compute time. But big-data workloads, due to highly random access nature, are not able to efficiently exploit the cache hierarchy. Consequently, they cause huge cache-flushes and data retrievals from off-chip memory~\cite{awan2015performance,awan2016micro}. This off-chip movement leads to substantial energy consumption and stalling of compute resources which causes performance degradation. Recent advancements in technology, however, has enabled us to bring the compute units in the close proximity of data. This new paradigm of bringing the computation closer to where the data reside is known as near-memory computing (NMC).

Big-data workloads, due to highly random access nature, are not able to efficiently exploit the cache hierarchy. Consequently, they cause huge cache-flushes and data retrievals from off-chip memory~\cite{awan2015performance,awan2016micro}. This off-chip movement leads to substantial energy consumption and stalling of compute resources, which causes performance degradation. Recent advancements in technology, however, have enabled us to bring the compute units in the proximity of data, which has lead to renewed interest in near-memory computing (NMC) architectures. Existing literature shows that these architecture are effective for a wide range of application ranging from graph processing to data management~\cite{8491877}. However, it's not trivial to identify those kernels which would benefit from the NMC paradigm and most of the studies rely on profiling applications of interest using e.g. hardware performance counters. The influence of micro-architecture features limits this technique.

To avoid this pitfall, we propose the platform-independent approach of characterizing workloads from NMC paradigm perspective. The idea is to profile instruction traces and collect inherent application information related to memory behavior  and parallelism. For that, we extend the capabilities of PISA~\cite{Anghel2016}, a platform-independent software analysis tool to extract characteristics directed towards NMC architectures~\cite{corda2019scopes} and we propose a method to validate the relevance of proposed metrics in selecting the kernels to offload on NMC hardware.

%Our contributions are:

%\begin{itemize}
%\item We present NMC specific micro-architecture independent application metrics.  
%\item We extend state the of the art platform-independent software analysis tool PISA with the NMC specific micro-architecture independent metrics~\cite{corda2019scopes}.
%\item We propose a method to validate the relevance of proposed metrics in selecting the kernels to offload on NMC hardware.
%\item We present prior art on workload characterization tools and position our work in the existing literature.
%\end{itemize}

\begin{comment}

The rest of the paper is organized as follows: \emph{Section \ref{sec:backgrounds}} presents the LLVM framework and the tool architecture that we extended.
%The characterization metrics that we embedded into the tool are described in \emph{Section \ref{sec:metrics}}.
In \emph{Section \ref{sec:metrics}} we describe the characterization metrics we embedded into PISA.
The methodology is presented in \emph{Section \ref{sec:methodology}}. In \emph{Section \ref{sec:results}} we show and discuss the characterization results. We give an overview of the main research on workload characterization in general and from the NMC perspective in particular in \emph{Section \ref{sec:relatedworks}}. Finally, \emph{Section \ref{sec:conclusions}} concludes this paper.

\end{comment}

%\raggedbottom

%% file: sections/back_pisa.tex
\begin{figure}[ht]
\centering
\includegraphics[width=8.5cm]{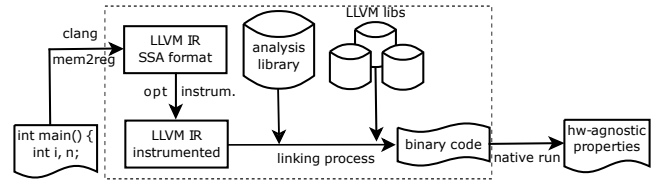}
\caption{Overview of the Platform-Independent Software Analysis Tool~\cite{Anghel2015AnIA}.
\label{fig:pisa}}
\end{figure}

PISA is based on the LLVM Compiler framework.
PISA's architecture is shown in \emph{Figure \ref{fig:pisa}}. Initially, the application source code, e.g., C/C++ code, is translated into the LLVM's intermediate representation (IR) using a clang front-end. This IR is independent of the target architecture and has a RISC-like instruction set.
PISA exploits the \emph{opt} tool to perform LLVM's IR optimizations and to perform the instrumentation process using an LLVM pass. This process is done by inserting calls to the external analysis library throughout the application's IR.
The last step consists of a linking process that generates a native executable. On running this executable, we can obtain analysis results for specified metrics in JSON format. PISA can extract metrics such as instruction mix, branch entropy, data reuse distance, etc. Moreover, PISA supports the MPI and OpenMP standards allowing the analysis of multi-threaded and multi-process applications.

%The tool's overhead depends on the analysis performed. On average the execution-time increases by two to three orders of magnitude in comparison to the non-instrumented code. However, since the analysis is target-independent, this has to be performed only once per application and dataset.

We extend PISA with metrics directed towards to NMC (PISA-NMC) \cite{corda2019scopes}.
%In this section we summarize (see \emph{Table \ref{tab:newmetrics}}) the metrics we integrated into PISA~\cite{corda2019scopes}. 
We focus on the memory behaviour, which is essential to decide if an application should be accelerated with NMC architecture, and on the parallelism behaviour, which is crucial to decide if a specific parallel architecture should be integrated into an NMC system.

\subsection{Memory metrics}
We add memory entropy, which measures the randomness of the memory addresses accessed using an entropy formula adapted to the memory accesses. If the memory entropy is high, which means a higher cache miss ratio, the application may benefit from 3D-stacked memory bandwith because of the volume of data moved from the main memory to the caches.

Data reuse distance or data temporal reuse (DTR) is a helpful metric to detect cache inefficiencies. The DTR of an address is the number of unique addresses accessed since the last reference of the requested data. We integrate this metric into PISA for different cache line sizes in order to compute the spatial locality metric.
Spatial locality measures the probability of accessing nearby memory locations.
The key idea behind this spatial locality score is to detect a reduction in DTR when doubling the cache line size. Usually, application with low spatial locality perform very bad on traditional systems with cache hierarchies because a small portion of data is utilized compared to the data loaded from the main memory to the caches

\subsection{Parallelism metrics}
Data-level parallelism (DLP) measures the average length of vector instructions that are used to optimize a program. DLP could be interesting for NMC when employing specific SIMD processing units in the logic layer of the 3D-stacked memory. We specialize the instruction-level parallelism (ILP) per opcode in order to estimate the DLP.

A basic-block is the smallest component in the LLVM's IR that can be considered as a potential parallelizable task. Basic-block level parallelism (BBLP) is a potential metric for NMC because it can estimate the task level parallelism in the application. The parallel tasks can be offloaded to multiple compute units located on the logic layer of a 3D-stacked memory. We develop metrics similar to ILP considering the basic block as a set of instruction that can be run only sequentially.

We also aim to estimate the presence of data parallel loops. Data parallel loops consist of basic-blocks that are repeated without any dependencies among their instances. We develop a metric, PBBLP (potential basic-block-level parallelism) that tries in a fast and straightforward manner to estimate the basic-block level parallelism in data-parallel loops.

%% file: sections/methodology.tex
To validate the relevance of our NMC specific metrics, we characterize the representative benchmarks using PISA-NMC and extract the numbers for memory entropy, spatial locality, data level parallelism, and basic-block level parallelism. We apply principal component analysis (PCA)~\cite{Jolliffe:1986}, on the collected metrics to make it more understandable by reducing its dimensionality~\cite{Hoste2006ComparingBU}. Next, we run the same benchmarks on an NMC system using a simulator and measure the improvement in energy-delay product and correlate this data with the output of PCA.  

\subsection{Host and NMC system}
\begin{figure}[t]
\centering
\includegraphics[scale=0.55]{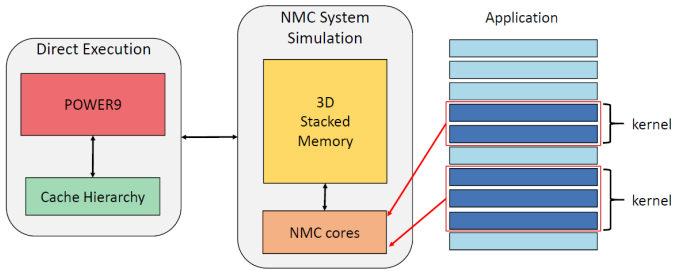}
\caption{Our NMC System
\label{fig:systemNMC}}
\end{figure}

Figure~\ref{fig:systemNMC} depicts the reference computing platform that we consider in this work. We run the applications both on a traditional Von-Neumann Architecture using the latest \emph{IBM Power 9}~\cite{power9} and on an NMC system based on hybrid memory cube (HMC). HMC memory is divided into several vertical DRAM partitions, called vaults, each with its own DRAM controller in the logic layer. In this work, we model NMC PEs as in-order, single-issue cores {with a private cache} as proposed in previous work~\cite{tesseract,7429299}. \emph{Table \ref{tab:systemchar}} lists the details of the host and NMC system used in our experiments. We extract the power consumption with AMESTER\footnote{https://github.com/open-power/amester} tool on Power 9. We simulate the NMC system on an extended version of the memory simulator \emph{Ramulator}~\cite{7063219} including the processing units. %\gaganRemove{The memory simulator performs system simulation by using traces of program execution.} 
Each processing unit is assigned to a vault and operates on the data assigned to that vault. We collect dynamic execution traces of the instrumented code with a Pin tool. We feed the acquired traces to Ramulator.

\begin{table}[ht]
\caption{Host and NMC System Characteristics}
\label{tab:systemchar}
\centering
\scalebox{0.98}{
\begin{tabular}{l|l|l|l}
\hline
\multicolumn{1}{c|}{\textbf{Architecture}} & \multicolumn{1}{c|}{\textbf{CPU Used}} & \multicolumn{1}{c|}{\textbf{Cache per core}} & \multicolumn{1}{c}{\textbf{Memory}} \\ \hline
\begin{tabular}[c]{@{}l@{}}IBM\\ Power9\\ (Host)\end{tabular} & \begin{tabular}[c]{@{}l@{}}4 cores \\ (SMT4)\\ @ 2.3 GHz\end{tabular} & \begin{tabular}[c]{@{}l@{}}L1 32 KB\\ L2 256 KB\\ L3 10 MB\end{tabular} & \begin{tabular}[c]{@{}l@{}}DDR4, 32 GB \\ RDIMM @\\ 2.7 GHz\end{tabular} \\ \hline
NMC & \begin{tabular}[c]{@{}l@{}}32 single-\\ issue in-order \\ cores @ \\ 1.25 GHz\end{tabular} & \begin{tabular}[c]{@{}l@{}}L1-I/D 2-way\\ 2 cache lines \\ 64B per cache \\ line\end{tabular} & \begin{tabular}[c]{@{}l@{}}HMC, 4GB\\ 8 stacked-layers,\\ 32 vaults, 16-bit\\ full duplex and \\ SerDes I/O link \\ @ 15 Gbps\end{tabular} \\ \hline
\end{tabular}
}
\end{table}

\subsection{Benchmarks}
Existing literature is devoid of proper benchmarks to evaluate NMC systems and explore the design space. Most of the studies instead design NMC systems tailored to improve the performance of specific workloads~\cite{overviewpaper}. We select a set of applications from two benchmark suites that are representative of the most common kernels and have been used previously in other related studies: Rodinia~\cite{zhang2014top,7551394,7756764} and Polybench~\cite{7756764}. Rodinia~\cite{5306797} is a benchmark suite for heterogeneous computing. Rodinia workloads cover a wide range of different behaviors which help a developer in building new systems. Polybench~\cite{pouchet2012polybench} is a collection of a large number of common kernels like matrix multiplication, stencil, covariance, correlation, etc. Each kernel is in a single file, tunable at compile-time. This makes instrumentation easier.

%% file: sections/results.tex
We present the NMC specific characterization of selected applications from PolyBench and Rodinia benchmark. Then, we show the correlation between the added metrics to NMC performance.

\subsection{Application characterization}
\begin{figure}[ht]
\centering

\includegraphics[width=8.5cm,trim={3cm 1cm 3cm 1cm},clip]{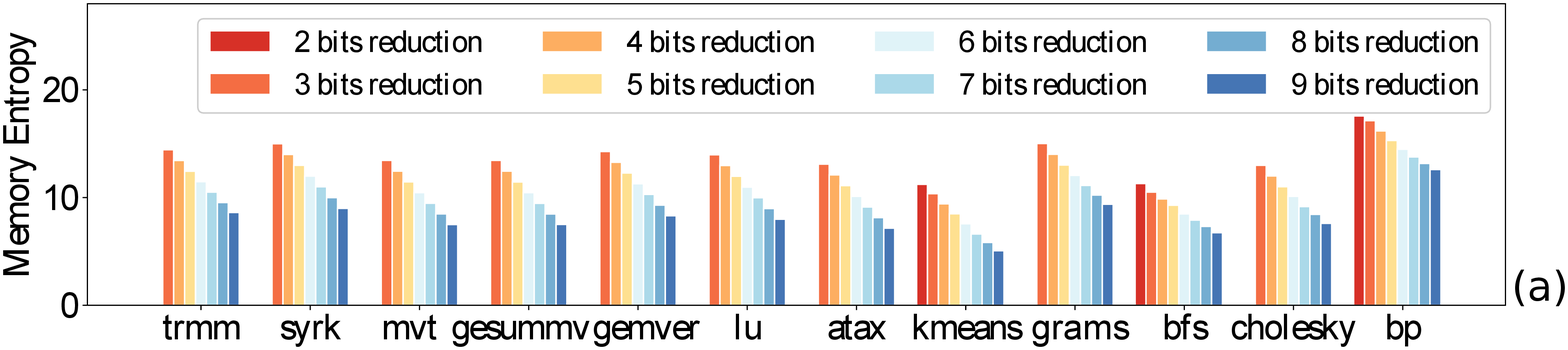}
\includegraphics[width=8.5cm,trim={3cm 1cm 3cm 1cm},clip]{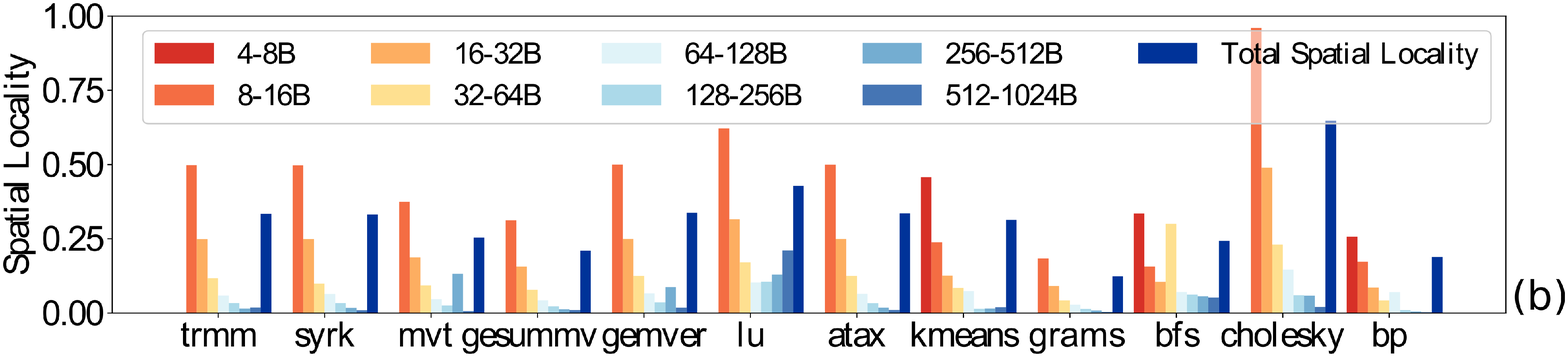}
\includegraphics[width=8.5cm,trim={3cm 1cm 3cm 1cm},clip]{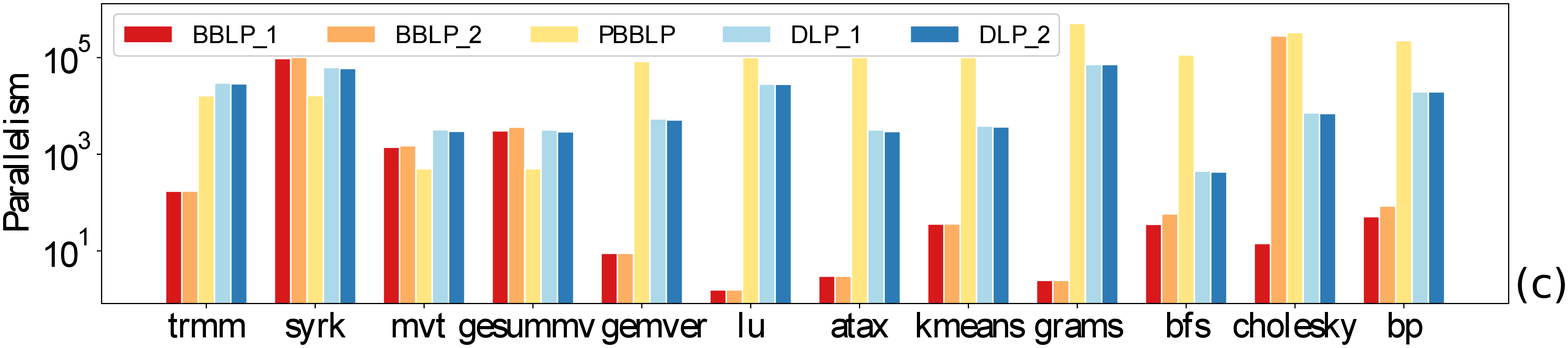}

\caption{Application characterization results:(a) Memory Entropy; (b) Spatial Locality; (c) Parallelism.} \label{fig:analysis}

\end{figure}

In \cite{corda2019scopes} we characterize a set of applications from Polybench and Rodinia for the added metrics (see \emph{Figure \ref{fig:analysis}}).  Memory entropy, in \emph{Figure \ref{fig:analysis}.a}, and spatial locality, in \emph{Figure \ref{fig:analysis}.b}, show respectively high and low values for applications such as \texttt{bp} and \texttt{gramschmidt} that could benefit from NMC because of their poor memory performance. \emph{Figure \ref{fig:analysis}.c} show the parallelism characterization of these kernels. In particular a group of applications shows good level of DLP, lowest level of BBLP and highest level of PBBLP. They could benefit from NMC architecture that exploit data-level parallel processing element.

%We characterize a set of applications from Polybench and Rodinia for the added metrics (see \emph{Figure \ref{fig:analysis}}). Memory entropy, in \emph{Figure \ref{fig:analysis}.a} is strictly related to the dimension of the address space accessed by a workload. The applications with larger address space have higher entropy because they are accessing many different addresses. We also plot memory entropy changes at different granularity cutting the least-significant bits (LSBs) of the address to represent larger data access granularity. Furthermore, we highlight in Rodinia's applications the cut of 2 LSBs because they are accessing integer (4Byte locations). We notice that applications like \texttt{bp} and \texttt{gramschmidt} have higher values of entropy and they should benefit from NMC architectures. Contrariwise, the other applications have similar values except for \texttt{cholesky}, \texttt{bfs} and \texttt{kmeans}. 

\subsection{Applications evaluation on NMC system}
We compare our metrics to the performance achieved by running the applications on an NMC system. We perform a single-thread analysis to estimate our metrics and then evaluate the execution time on the considered architectures. Table~\ref{tab:results} lists the parameter levels for the evaluated applications on the Power 9 and NMC system. We consider only single-thread analysis here to avoid the side effects of a multi-threaded analysis in metrics such as memory entropy and spatial locality, e.g., averaging the numbers from multiple threads tend to mask the true behavior of applications. Since the analysis trend is similar for different dataset sizes and the memory analysis is highly time-consuming we use smaller dataset than the one simulated for the NMC system in line with similar work on application characterization~\cite{Anghel2016, Gu:2009:CMS:1542431.1542446}. \emph{Figure \ref{fig:edp}} shows the energy-delay product (EDP) ratio between the IBM Power 9 and the NMC system we simulated. We use EDP as our major metric of reference in this analysis because both energy and performance are critical criteria for evaluating NMC suitability. Applications with EDP reduction less than 1 are not suitable for NMC.  

\begin{table}[H]
%\caption{Application parameters used for the evaluation on NMC and Host system}
\caption{Benchmarks Parameters}

\label{tab:results}
\centering
\scalebox{0.98}{
\begin{tabular}{l|l|ll}
\hline
\multicolumn{2}{c|}{\textbf{Applications}} & \multicolumn{2}{c}{\textbf{Parameters}} \\ \hline
\multicolumn{1}{c|}{\textbf{Benchmarks}} & \multicolumn{1}{c|}{\textbf{Kernels}} & \multicolumn{1}{c|}{\textbf{Param.}} & \multicolumn{1}{c}{\textbf{Values}} \\ \hline
\multirow{2}{*}{Polybench} & atax, gemver, gesummv & \multicolumn{1}{l|}{dimensions} & 8000 \\ \cline{2-4} 
 & \begin{tabular}[c]{@{}l@{}}cholesky, gramschmidt, \\ lu, mvt, syrk, trmm\end{tabular} & \multicolumn{1}{l|}{dimensions} & 2000 \\ \hline
\multirow{3}{*}{Rodinia} & bfs & \multicolumn{1}{l|}{nodes} & 1.0m \\ \cline{2-4} 
 & bp & \multicolumn{1}{l|}{layer size} & 1.1m \\ \cline{2-4} 
 & kmeans & \multicolumn{1}{l|}{data size} & 819k \\ \hline
\end{tabular}
}
\end{table}

\subsection{Correlation between NMC Metrics and NMC Performance}
\begin{figure}[H]
\centering
\includegraphics[scale=0.55]{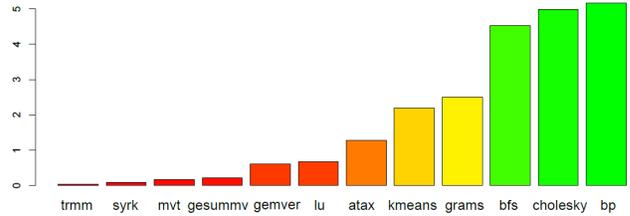}
\caption{EDP improvement
\label{fig:edp}}
\end{figure}

Spatial locality in \emph{Figure \ref{fig:analysis}.b} provides insights on which application could be better for the NMC system we considered. Applications that show the lowest spatial locality such as \texttt{gramschmidt}, \texttt{bp}, \texttt{bfs} show a considerable EDP improvement (see \emph{Figure \ref{fig:edp}}) using the NMC system. Contrariwise also \texttt{cholesky}, that has the highest spatial locality among the chosen applications, benefits from the NMC architecture.
Memory entropy in \emph{Figure \ref{fig:analysis}.a} gives similar insights. For instance, applications with the highest entropy such as \texttt{gramschmidth} and \texttt{bp} shows benefit executing on an NMC system. However, also applications with low entropy seem to benefit from NMC. 
Parallelism analysis in \emph{Figure \ref{fig:analysis}.c} highlights that most of the applications that benefit from NMC have the lowest values for $BBLP_1$ and a good level of DLP. However, there are some exceptions such as \texttt{lu}, that has the lowest BBLP values, and \texttt{bfs} that has the lowest DLP values.
The above shows that  a single metric can not explain NMC appropriateness. To get more insights into what combinations of metrics can predict NMC applicability, we apply PCA to our metric results. %For this we include two metrics derived from already existing features in PISA: Instruction Mix and ILP (see \emph{Figure \ref{fig:instmix}}). First, we evaluate the ratio of memory instructions (loads and stores) and arithmetic instructions (floating point and integer) that can be helpful using a NMC architecture where memory operations are more relevant. Then, as another parallelism indicator we select the ILP normalized by the number of instructions to make the comparison of the average ILP easier among different applications.
%\textbf{

\begin{figure}[ht]
\centering
\includegraphics[scale=0.2]{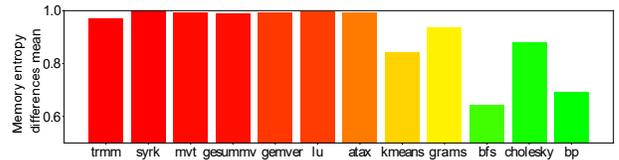}
\caption{Metric derived from memory entropy.
\label{fig:entropydiff}}
\end{figure}

For this, we derive another metric from the memory entropy exploiting the granularity. For each application, we first compute the difference between each couple of consecutive memory entropy values with different granularities (see \emph{Figure \ref{fig:analysis}.a}, larger granularity represents larger cache line size). Then, we compute the average of these values that represents a spatial locality variation increasing the cache line size. \emph{Figure \ref{fig:entropydiff}} shows this metric. This metric compared to the EDP values shows that the major part of the applications not suitable for NMC has the highest values.
%}

\begin{figure}[ht]
\centering
\includegraphics[scale=0.25]{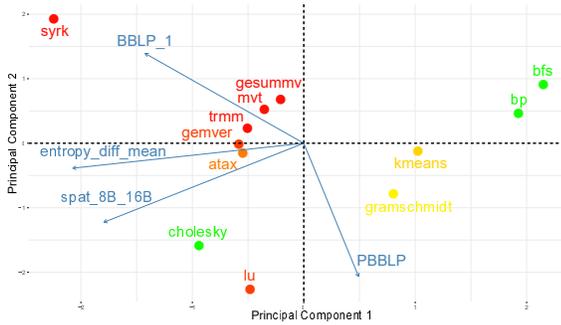}
\caption{PCA using the added metrics. Blue arrows quantify the contribution and direction to the PCs.% The axis are the two most dominant principal components. 
\label{fig:pca}}
\end{figure}

 \emph{Figure \ref{fig:pca}} shows the PCA applied to the most promising subset of presented metrics. We use 4 input features for the PCA: $BBLP_1$, $PBBLP$, $entropy\_diff\_mem$ (the value proposed above) and $spat\_8B\_16B$ (spatial locality doubling the cache line size from 8B to 16B).
 We highlight that all the applications that benefit from Power9 are in the II quadrant (top-left) except for \texttt{lu} that is in the III quadrant. In its code diagonal matrix accesses are present and they should be critical for traditional CPUs. It could be an NMC application candidate employing a larger dataset size.
 The applications that benefits from NMC are in the other quadrants. In particular \texttt{bfs} and \texttt{bp} seem having similar characteristic and are located in the I quadrant. Similarly \texttt{gramschmidt} and \texttt{kmeans} located in the IV quadrant.
 These metrics show good potential in discriminating NMC applications.

%% file: sections/relatedworks.tex
Existing studies primarily rely on hardware performance counters available in the modern processors to understand the memory access behavior of the applications and identify the kernels suitable for offload to NMC architectures~\cite{awan2017identifying, ahn2015pim, awan2017performance,Boroumand:2018:GWC:3296957.3173177,zhang2014top}. Others have used a dynamic binary instrumentation framework like Pin~\cite{7284059} or estimation at the compile time~\cite{7756764,7551394} for the same purpose. PISA-NMC showed a different approach to workload characterization applied to NMC. We used a target-agnostic workload characterization technique to extract metrics directed towards NMC. Then, we used PCA and NMC simulation to show the relevance of the metrics proposed.

%% file: sections/conclusions.tex
%Near-memory computing (NMC) is still in its first stages of development and therefore lacks proper tools for specialized workload profiling. Application characterization from NMC perspective is crucial in selecting and subsequently scheduling the NMC kernels on the NMC system.
%We have extended PISA, a state-of-the-art application characterization tool, with metrics relevant for NMC architectures. Particularly, we have concentrated on analyzing the memory accesses and parallelism behaviors: data-level parallelism, basic-block level parallelism, memory entropy, and spatial locality.

%We have highlighted the importance of the presented metrics for an NMC system by correlating the NMC specific application characteristics with the application performance on an NMC system.
%As expected, memory behavior is crucial to detect application suitable for NMC. Indeed we have shown the importance of the memory entropy and spatial locality. Furthermore we also have highlighted how parallelism helps in discriminating the NMC applications. 
%However, we observed that there are different behaviors across NMC applications and that multiple metrics are needed to explain suitability for NMC.

We extend PISA with NMC specific metrics such as data-level parallelism, basic-block level parallelism, memory entropy,  and spatial locality.  By  correlating  the  principal  components  of metrics as mentioned above  with  the  energy-delay product of benchmark kernels on an NMC system, we show that PISA-NMC  can  help to identify the  kernels  that  can  benefit from NMC in a platform agnostic manner.
As future work, we will investigate more workloads and perform a more exhaustive analysis.

%As expected, applications with the lowest and highest values for spatial locality perform better on NMC. Furthermore, some of the parallelism metrics give interesting insights in discriminating NMC applications.
%However, we observed that there are different behaviors across NMC applications and that multiple metrics are needed to explain suitability for NMC.

%In the future, the tool can be used to perform an NMC performance prediction based on the metrics. Another interesting future challenge would be to collect different applications in NMC benchmark suite.

%\color{blue} We highlighted the benefit of these metrics for NMC architectures, for instance, the high importance of parallelism and spatial locality. Then, we showed which applications could be potential candidates for NMC, such as matrix multiplication and stencil algorithms that show high-level of parallelism with a fair amount of spatial locality. We also demonstrated the multi-threaded capabilities of PISA for the added metrics, characterizing a set of Rodinia applications.
%\color{black}

%In the future, the tool could be extended to perform an analysis focused on other emerging architectures. Additionally, this enhanced version of PISA could be used for design space exploration (DSE) or application classification. 

%% file: sections/acks.tex
This work is funded by European Commission under Marie Sklodowska-Curie Innovative Training Networks European Industrial Doctorate (Project ID: 676240). 